\documentclass[oneside,english]{siamltex}
\usepackage[T1]{fontenc}
\usepackage[latin9]{inputenc}
\usepackage{listings}
\usepackage{xcolor}
\usepackage{float}
\usepackage{amsbsy}
\usepackage{amstext}
\usepackage{amssymb}
\usepackage{graphicx}
\usepackage{esint}
\PassOptionsToPackage{normalem}{ulem}
\usepackage{ulem}

\makeatletter

\providecommand{\tabularnewline}{\\}
\providecolor{lyxadded}{rgb}{0,0,1}
\providecolor{lyxdeleted}{rgb}{1,0,0}
\newcommand\eqref[1]{(\ref{#1})}

\usepackage{tikz}
\usetikzlibrary{arrows,decorations.pathreplacing}

\usepackage{listings}
\usepackage{mathabx}
\usepackage{relsize}

\@ifundefined{showcaptionsetup}{}{%
 \PassOptionsToPackage{caption=false}{subfig}}
\usepackage{subfig}
\makeatother

\usepackage{babel}
\begin{document}

\title{Generalized sensitivity functions for size-structured population
models}

\author{Dustin D. Keck\thanks{Applied Mathematics, University of Colorado, Boulder, CO 80309-0526}
\and David M. Bortz$^{*}$\thanks{Corresponding author (dmbortz@colorado.edu)}}
\maketitle
\begin{abstract}
Size-structured population models provide a popular means to mathematically
describe phenomena such as bacterial aggregation, schooling fish,
and planetesimal evolution. For parameter estimation, generalized
sensitivity functions (GSFs) provide a tool that quantifies the impact
of data from specific regions of the experimental domain. These functions
help identify the most relevant data subdomains, which enhances the
optimization of experimental design. To our knowledge, GSFs have not
been used in the partial differential equation (PDE) realm, so we
provide a novel PDE extension of the discrete and continuous ordinary
differential equation (ODE) concepts of Thomaseth and Cobelli and
Banks et al.~respectively. We analyze the GSFs in the context of
size-structured population models, and specifically analyze the Smoluchowski
coagulation equation to determine the most relevant time and volume
domains for three, distinct aggregation kernels. Finally, we provide
evidence that parameter estimation for the Smoluchowski coagulation
equation does not require post-gelation data.
\end{abstract}

\section{Introduction\label{sec:Introduction}}

General structured population models provide a link from the individuals
in a population to the population processes \cite{Easterling2000,Ebenman1988,Tuljapurkar1997}.
A popular example, size-structured population models describe the
distribution of individuals throughout varying size classes \cite{Cushing1987,Doumic2010a}.
Typical ODE based population models make a number of simplifying assumptions,
a major one of which presumes homogeneity of the individuals' physical
structure across the entire population. One effort to relax the homogeneity
assumption resulted in the creation of age-structured population models
which account for the effects of differing ages amongst the individuals
comprising the population. Unfortunately, for some systems, age does
not comprise the most influential physical attribute, but in many
of these cases, size-structured population models do provide an adequate
structuring of the population \cite{Cushing1990}.

Size-structured population models often include an unknown parameter
(or a set of unknown parameters). The value of this parameter is estimated
via the inverse problem of parameter estimation based on experimental
data. With a goal of optimizing the experiments, we seek to sample
from domains which contain the most relevant information regarding
the parameter estimation. Generalized sensitivity functions provide
a tool which quantifies the importance of specific regions of a domain
to the parameter of interest. Previous studies, such as cardiovascular
regulation \cite{Batzel2009,Fink2006,Fink2008}, HIV modeling \cite{DavidTranBanks2009},
and HTLV-1 transactivation simulation \cite{Corradin2011}, have applied
the generalized sensitivity functions to ordinary differential equations.
We denote these ODE-based GSFs as OGSFs. With our emphasis on size-structured
population models, the primary goal of our work is to extend the concepts
of OGSFs to the application of generalized sensitivity functions to
PDEs, which we denote as PGSFs.

Thomaseth and Cobelli introduced the concept of OGSFs in \cite{Thomaseth1999}%
\footnote{Note that in the original Thomaseth and Cobelli work, the functions
are simply called GSFs (not OGSFs), since the authors are only considering
ODE-based models.%
} andBatzel et al.~recast the OGSFs into a probabilistic setting \cite{Batzel2007a}.
In a series of studies, Banks et al.~\cite{BanksDavidianSamuelsSutton2009,Banks2010,BanksDediuErnstberger2008}
further develop the OGSF concept.. In particular, the work by Banks,
Dediu, and Ernstberger \cite{BanksDediuErnstberger2008} compares
traditional sensitivity functions (TSFs) with OGSFs (in the context
of general nonlinear ODEs) and highlights the potential utilities
of OGSFs. In \cite{BanksDediuErnstberger2008}, the authors also warn
that OGSFs possess a potential weakness, which they denote as the
\textit{forced-to-one artifact }(discussed in Section \ref{sec:GSF-Theory})\textit{.}
Then Banks, Davidian, Samuels, and Sutton \cite{BanksDavidianSamuelsSutton2009}
expand the results in \cite{BanksDediuErnstberger2008} by introducing
methodology for choosing between TSFs and OGSFs. Later, Banks, Dediu,
Ernstberger, and Kappel \cite{Banks2010} extend the OGSFs to a continuous
setting and demonstrate the value of the OGSFs in the context of optimal
experimental design.

As a case study for our extension of OGSFs to the PDE context, we
apply our PGSFs to the Smoluchowski coagulation equation. This model
for size-structured populations arises in the study of organic phenomena
such as bacterial growth \cite{BortzEtal2008bmb}, marine snow \cite{Kiorboe2001},
algal blooms \cite{Ackleh1997NATMA,AcklehFitzpatrick1997,Riebesell1992},
and schooling fish \cite{Niwa1998} and inorganic phenomena such as
powder metallurgy \cite{Kumar2009}, astronomy \cite{Lee2000icarus,Lee2001jopa,Makino1998,Silk1978},
aerosols \cite{Drake1972}, irradiation of metals \cite{Surh2008},
and meteorology \cite{Pruppacher1980}. For our study, we determine
the time and volume subdomains, which we denote $\mathcal{D}^{*}$,
of greatest relevance to the estimate of the constant parameter in
three coagulation kernels. In Section \ref{sec:GSF-Theory}, we summarize
the original work on OGSFs and the extensions to it. We then make
a further extention of OGSFs to PGSFs for implementation on size-structured
population models. In Section \ref{sec:Application}, we discuss the
details of how we implement the PGSFs with respect to the Smoluchowski
coagulation equation. In Section \ref{sec:Results}, we provide our
results for each of three coagulation kernels. Finally in Section
\ref{sec:Conclusions-and-Future}, we summarize the conclusions we
have drawn from this study and discuss future directions for this
research.

\section{Generalized sensitivity functions Theory\label{sec:GSF-Theory}}

Given a domain $\mathcal{D}$ for the independent variables, the PGSFs
will allow us to identify a subregion $\mathcal{D}^{*}\subset\mathcal{D}$,
containing the information necessary to make the most accurate parameter
estimates. The OGSFs and PGSFs vary from the TSFs%
\footnote{For a summary of TSFs, see Stanley and Stewart \cite{Stanley2002}.%
} in the sense that the OGSFs and PGSFs do not depend on specific data
realizations, which we explain in more detail in Section \ref{sub:Discrete1D}.
When Thomaseth and Cobelli introduced the OGSFs in \cite{Thomaseth1999},
they argued that the subdomain over which the OGSFs most rapidly increase
to one contains the most relevant information for the parameter of
interest. Then in \cite{Banks2010}, Banks et al.~provide evidence
that subdomains over which the OGSFs most rapidly decrease (indicating
a high correlation between multiple parameters) also contain high
information content. 

In adddition to the OGSFs, Thomaseth and Cobelli provide a related
tool, the incremental (O)GSF, which computes the information at a
given time point informing the value of a parameter estimate \cite{Thomaseth1999}.
As advocated by the authors, the OGSFs and the incremental OGSFs should
be regarded as complementary to one another. To demonstrate the complementary
characteristics of OGSFs and incremental OGSFs, Thomaseth and Cobelli
present an example where the plots of the OGSFs suggest an optimal
$\mathcal{D}^{*}$. Banks et al.~define a related quantity, the time
derivative of the OGSFs, which plays the role of an incremental GSF
when the OGSF is defined over continuous time (see a similarly complementary
role to the continuous OGSFs. 

As mentioned in Section \ref{sec:Introduction}, one weakness of generalized
sensitivity functions is the so called \textit{forced to one artifact}
(FTOA). As addressed at length by Banks, Dediu, and Enrstberger in
\cite{BanksDediuErnstberger2008}, plots of the OGSFs vary with changes
in $\mathcal{D}$. Regardless of the choice of domain, by definition,
the OGSFs and PGSFs will attain a value of one at the independent
variables' maximum values in $\mathcal{D}$. Therefore, if $\mathcal{D}$
possesses insufficient maximum values, the generalized sensitivity
functions may provide misleading information about $\mathcal{D}^{*}$
because they were (by definition) forced to a value of one on the
upper bound of the domain. A strategy to counter this weakness \cite{Banks2010}
is to check that the time derivative of the OGSFs approaches zero
within the original choice of $\mathcal{D}$. If it does not, we extend
$\mathcal{D}$ untilthe derivative does satisfy this criteria.

In Section \ref{sub:Discrete1D}, we summarize Thomaseth's and Cobelli's
and Banks et al.'s development of the discrete and continuous OGSFs,
respectively. In Section \ref{sub:Cont2D GSF}, we extend these previous
works to the continuous PGSFs setting necessary for parameter estimation
in size-structured population, PDE models. Finally, in Section \ref{sub:subdomain_determ},
we propose mathematical criteria for determining $\mathcal{D}^{*}$.

\subsection{ODE-Based GSFs (OGSF)\label{sub:Discrete1D}}

In this section, we summarize the theory introduced by Thomaseth and
Cobelli in \cite{Thomaseth1999} and Banks et al.~in \cite{BanksDavidianSamuelsSutton2009,Banks2010,BanksDediuErnstberger2008}.
We provide this summary as convenient setting for introducing much
of the notation and many of the definitions needed throughout this
work.

First, we represent the system under consideration as a nonlinear
regression function $f(t;\theta)$ with $t$ representing the sole
independent variable and with $\boldsymbol{\theta}=[\theta_{1},\theta_{2},\dots,\theta_{L}]^{T}$
representing the parameter column vector with dimension $L$.%
\footnote{ Note that bold typeface indicates a vector quantity.%
} Then we represent the measurements with noise as
\begin{equation}
y(t)=f(t;\boldsymbol{\theta})+\epsilon(t),\,\label{eq:noisy observation}
\end{equation}
where $\epsilon(t)$ is the measurement noise. We assume an independent
identically distributed noise distribution with zero mean and with
known (but possibly time varying) variance, $\sigma^{2}(t)$. We also
assume the existence of a true parameter vector $\boldsymbol{\theta}_{0}$.
When the observation times are discrete (as in \cite{Thomaseth1999}),
the generalized sensitivity is defined as 
\begin{eqnarray}
\boldsymbol{gs}(t_{k}) & = & \sum_{i=1}^{k}\left\{ \left(\left[\sum_{j=1}^{N_{t}}\frac{1}{\sigma^{2}(t_{j})}\nabla_{\boldsymbol{\theta}}f(t_{j};\boldsymbol{\theta}_{0})\nabla_{\boldsymbol{\theta}}f(t_{j};\boldsymbol{\theta}_{0})^{T}\right]^{-1}\frac{\nabla_{\boldsymbol{\theta}}f(t_{i};\boldsymbol{\theta}_{0})}{\sigma^{2}(t_{i})}\right)\right.\nonumber \\
 & \bullet & \nabla_{\boldsymbol{\theta}}f(t_{i};\boldsymbol{\theta}_{0})\Biggr\},\label{eq:discreteGSF}
\end{eqnarray}
where $\bullet$ indicates a Hadamard product and $N_{t}$ is the
number of timepoints.In the appendix to \cite{Thomaseth1999}, the
authors also introduce the incremental OGSFs defined as
\[
\boldsymbol{gs}_{inc}(t_{k})=\boldsymbol{gs}(t_{k})-\boldsymbol{gs}(t_{k-1}),
\]
yielding

\begin{eqnarray}
\boldsymbol{gs}_{inc}(t_{k}) & = & \left(\left[\sum_{j=1}^{N_{t}}\frac{1}{\sigma^{2}(t_{j})}\nabla_{\boldsymbol{\theta}}f(t_{j};\boldsymbol{\theta}_{0})\nabla_{\boldsymbol{\theta}}f(t_{j};\boldsymbol{\theta}_{0})^{T}\right]^{-1}\frac{\nabla_{\boldsymbol{\theta}}f(t_{k};\boldsymbol{\theta}_{0})}{\sigma^{2}(t_{k})}\right)\nonumber \\
 & \bullet & \nabla_{\boldsymbol{\theta}}f(t_{k};\boldsymbol{\theta}_{0}).\label{eq:gsinc}
\end{eqnarray}
With this definition (\ref{eq:gsinc}), one can calculate the contribution
of the partial derivative at a specific point, $t_{k}$, rather than
sum all contributions at times up to and including $t_{k}.$

Banks et al.~developed a continuous version of the generalized sensitivity
functions
\begin{equation}
gs(t;\boldsymbol{\theta})=\int_{0}^{t}\left(F(\overline{t};\boldsymbol{\theta})^{-1}\frac{1}{\sigma^{2}(s)}\nabla_{\boldsymbol{\theta}}f(s;\boldsymbol{\theta}_{0})\right)\bullet\nabla_{\boldsymbol{\theta}}f(s;\boldsymbol{\theta}_{0})dP(s),\,\,\, t\in[0,\overline{t}].\label{eq:contGSF}
\end{equation}
where $F(\overline{t};\boldsymbol{\theta})$ represents the generalized
Fisher information matrix \cite{Banks2011} 
\[
F(\overline{t};\boldsymbol{\theta}):=\int_{0}^{\overline{t}}\frac{1}{\sigma^{2}(s)}\nabla_{\boldsymbol{\theta}}f(s;\boldsymbol{\theta})\nabla_{\boldsymbol{\theta}}f(s;\boldsymbol{\theta})^{T}dP(s)\,.
\]
As a tool to prevent misleading conclusions from a potential FTOA,
Banks et al\@.~futher introduce the time derivative of $gs(t;\theta)$
\begin{equation}
\frac{\partial}{\partial t}gs(t;\boldsymbol{\theta})\coloneq\left(F(\overline{t};\boldsymbol{\theta})^{-1}\frac{1}{\sigma^{2}(s)}\nabla_{\boldsymbol{\theta}}f(s;\boldsymbol{\theta}_{0})\right)\bullet\nabla_{\boldsymbol{\theta}}f(s;\boldsymbol{\theta}_{0}).\label{eq:contincGSF}
\end{equation}
In our subsequent work, we denote this quanity (\ref{eq:contincGSF})
as $gs_{\mathsmaller{RIA}}(t;\boldsymbol{\theta})=\frac{\partial}{\partial t}gs(t;\theta)$,
i.e., the \textit{rate of information acquisition} (RIA) at a specific
point in $\mathcal{D}$.

\subsection{Continuous PDE-Based GSFs (PGSF)\label{sub:Cont2D GSF}}

In the OGSFs studies, the nonlinear regression function, $f(t;\boldsymbol{\theta})$,
contains one independent variable, and a vector of parameters. For
general size-structured population, continuous PDE models, we adapt
the nonlinear regression function to depended on a column vector of
independent variables, which we denote as $\mathbf{r}=[r_{1},r_{2,}\dots,r_{N_{r}}]^{T}$
with dimension $N_{r}$. For example, in the analysis of the OGSFs
in Section \ref{sub:Discrete1D}, $N_{r}=1$ and $r_{1}=t$, whereas
with the Smoluchowski coagulation PDE, $N_{r}=2$ and $[r_{1},r_{2}]=[t,x]$.
Without loss of generality, we also let $r_{i}\in[0,\overline{r}_{i}]$
for each $i\in[1,2,\dots,N_{r}]$, where $\overline{r}_{i}$ represents
the maximum values of each independent variable respectively, and
we denote the vector of maximum independent variable values $\overline{\mathbf{r}}=[\overline{r}_{1},\overline{r}_{2},\dots,\overline{r}_{N_{r}}]^{T}$.
From the continuous OGSFs defined by Banks et al.~in \cite{Banks2010},
we can then make the straight forward extension for the Fisher information
matrix
\[
F(\overline{\mathbf{r}};\boldsymbol{\theta}):=\int_{0}^{\overline{r}_{N_{r}}}\int_{0}^{\overline{r}_{N_{r}-1}}\cdots\int_{0}^{\overline{r}_{1}}\frac{1}{\sigma^{2}(\mathbf{r})}\nabla_{\boldsymbol{\theta}}f(\boldsymbol{\mathtt{r}};\boldsymbol{\theta})\nabla_{\boldsymbol{\theta}}f(\mathbf{r};\boldsymbol{\theta})^{T}dr_{1}dr_{2}\dots dr_{N_{r}}\,,
\]
and a continuous PGSF, 
\begin{eqnarray*}
gs(\mathbf{r};\boldsymbol{\theta}) & = & \int_{0}^{r_{N_{r}}}\int_{0}^{r_{N_{r}-1}}\cdots\int_{0}^{r_{1}}\frac{1}{\sigma^{2}(\mathbf{r})}\left(F(\overline{\mathbf{r}};\boldsymbol{\theta})^{-1}\frac{1}{\sigma^{2}(\mathbf{r})}\nabla_{\boldsymbol{\theta}}f(\mathbf{r};\boldsymbol{\theta}_{0})\right)\\
 & \bullet & \nabla_{\boldsymbol{\theta}}f(\mathbf{r};\boldsymbol{\theta}_{0})dr_{1}dr_{2}\dots dr_{N_{r}}\,\,\,\,\boldsymbol{\theta}\in\mathbb{R}^{L}\,.
\end{eqnarray*}
We also extend (\ref{eq:contincGSF}) to a \emph{rate of information
acquisition} (RIA) for a PGSF
\begin{eqnarray*}
gs_{\mathsmaller{RIA}}(\mathbf{r};\boldsymbol{\theta}) & = & \frac{\partial^{N_{r}}}{\partial r_{N_{r}}\partial r_{N_{r}-1}\cdots\partial r_{1}}gs(\mathbf{r};\boldsymbol{\theta})\coloneq\left(F(\overline{\mathbf{r}};\boldsymbol{\theta})^{-1}\frac{1}{\sigma^{2}(\mathbf{r})}\nabla_{\boldsymbol{\theta}}f(\mathbf{r};\boldsymbol{\theta}_{0})\right)\\
 & \bullet & \nabla_{\boldsymbol{\theta}}f(\mathbf{r};\boldsymbol{\theta}_{0}).
\end{eqnarray*}

In this work, we examine size-structured populations in a general
context, therefore we assume a constant variance of one and normal
error distribution for all measurements. Furthermore, for our purposes,
we adapt the nonlinear regression function so that $f$ depends on
two variables, $t$ and $x$, and one parameter, $\alpha$, so that
\begin{equation}
gs(t,x;\alpha)=\frac{\int_{0}^{t}\int_{0}^{x}\left(\frac{\partial f}{\partial\alpha}\right)^{2}drds}{\int_{0}^{\overline{t}}\int_{0}^{\overline{x}}\left(\frac{\partial f}{\partial\alpha}\right)^{2}drds},\label{eq:2DcontGSF}
\end{equation}
where $\overline{t}$ and $\overline{x}$ represents the maximum values
of time and volume respectively. Also, for the rate of information
acquisition,
\begin{eqnarray}
gs_{\mathsmaller{RIA}}(t,x;\alpha)\coloneq\frac{\partial^{2}}{\partial t\partial x}\left[gs(t,x;\alpha)\right] & = & \frac{\partial^{2}}{\partial t\partial x}\left[\frac{\int_{0}^{t}\int_{0}^{x}\left(\frac{\partial f}{\partial\alpha}\right)^{2}drds}{gs(\overline{t},\overline{x};\alpha)}\right]\nonumber \\
 & = & \frac{\left(\frac{\partial f(t,x;\alpha)}{\partial\alpha}\right)^{2}}{gs(\overline{t},\overline{x};\alpha)}.\label{eq:2Dcontgensensiar}
\end{eqnarray}

\subsection{Determining the optimum subdomain, $\mathcal{D}^{*}$\label{sub:subdomain_determ}}

In \cite{Thomaseth1999}, Thomaseth and Cobelli only offer a visual
analysis of how we can apply the OGSFs to determine $\mathcal{D}^{*}$.
In \cite{Banks2010}, Banks et al.~propose a mathematical implementation
to determine the upper bound of $\mathcal{D}^{*}$ by bounding the
TSFs variation. In this section, we offer a quantitative means for
computing the lower and upper bounds of an optimal $\mathcal{D}^{*}$.
To determine the lower ends of $\mathcal{D}^{*}$, we consider a level
curve that represents a fraction of the maximum RIA. Then to determine
the upper ends of $\mathcal{D}^{*}$, we consider a second level curve
that represents the points where the PGSFs approach a value of one.

First, we consider the RIA to determine the lower ends of $\mathcal{D}^{*}$.
In the analysis that follows, we assume only two independent variables
and one parameter with $\mathcal{D}=[\underbar{\ensuremath{t}},\overline{t}]\times[\underbar{\ensuremath{x}},\overline{x}]$.%
\footnote{The inclusion of additional independent variables and parameters is
straightforward.%
} We define the maximum $gs_{\mathsmaller{RIA}}$ as 
\[
\overline{gs}_{\mathsmaller{RIA}}=\max_{(t,x)\in\mathcal{D}}\{\left|gs_{\mathsmaller{RIA}}(t,x;\alpha)\right|\}.
\]
Then we denote a fraction $\gamma\in(0,1)$ of $\overline{gs}_{\mathsmaller{RIA}}$and
define the level curve, $\Gamma_{\gamma}$, where 
\[
\Gamma_{\gamma}=\left\{ (t,x)\left|\right.\left|gs_{\mathsmaller{RIA}}(t,x;\alpha)\right|=\gamma\overline{gs}_{\mathsmaller{RIA}}\right\} ,
\]
and from that level curve, we find the minimum values of $x$ and
$t$, which we denote $x_{*}$ and $t_{*}$, where $x_{*}=\min_{x}\Gamma_{\gamma}$
and $t_{*}=\min_{t}\Gamma_{\gamma}.$ 

Second, we consider the PGSFs to determine the upper ends of $\mathcal{D}^{*}$.
We let $\rho$ represent the proximity to one that we desire, and
define the level curve, $\Gamma^{\rho}$, where
\[
\Gamma^{\rho}=\left\{ (t,x)\left|\right.\left|1-gs(t,x;\alpha)\right|=\rho\right\} .
\]
Note that any point in the set $\Gamma^{\rho}$ provides a satisfactory
upper bound on $\mathcal{D}^{*}$. To determine a range of upper bounds
depending on which independent variable costs more in terms of gathering
data, we consider $x^{*}=\min_{x}\Gamma^{\rho}$ with its dependent
$t(x^{*})$, and $t^{*}=\min_{t}\Gamma^{\rho}$ with its dependent
$x(t^{*})$. We denote the optimum subdomain as $\mathcal{D}_{x}^{*}$,
where 
\[
\mathcal{D}_{x}^{*}=[t_{*},t(x^{*})]\times[x_{*},x^{*}],
\]
when high resolution data in $x$ is more expensive. Conversely, when
high resolution data in $t$ is more expensive, we denote the optimum
subdomain as $\mathcal{D}_{t}^{*}$, where
\[
\mathcal{D}_{t}^{*}=[t_{*},t^{*}]\times[x_{*},x(t^{*})].
\]

\section{Application of PGSFs to the Smoluchowski coagulation equation\label{sec:Application}}

Our extension to PGSFs allows us to apply it to the Smoluchowski coagulation
equation with one parameter of interest, the aggregation kernel constant.
In the early 1900's, van Smoluchowski developed a model to study the
coagulation of colloids \cite{Smoluchowski1916,Smoluchowski1917},

\begin{equation}
\frac{d}{dt}f_{k}=\frac{1}{2}\sum_{i+j=k}K(i,j)f_{i}f_{j}-\sum_{i}K(i,k)f_{i}f_{k},
\end{equation}
where $f_{k}$ represents the number density of aggregates of volume
$k$, and $K(i,j)$ is the aggregation kernel denoting the rate at
which aggregates of size $i$ and $j$ form a combined aggregate of
size $i+j$ \cite{BortzEtal2008bmb,Smoluchowski1916,Smoluchowski1917}.
Müller subsequently extended this model to a continuous PDE \cite{Filbet2004,Muller1928} 

\begin{eqnarray}
\partial_{t}f & = & A(f),\,\,(t,x)\in\mathbb{R}_{+}^{2},\label{eq:smolu}\\
f(0) & = & f_{0},\,\, x\in\mathbb{R}_{+}\nonumber 
\end{eqnarray}
 where each aggregate is classified solely by its volume $x>0$, and
$f=f(t,\cdot)$ represents the continuous size distribution function
of aggregates at time $t\ge0$. The coagulation term is

\begin{eqnarray}
A(f) & = & A_{in}(f)-A_{out}(f)\nonumber \\
 & = & \frac{1}{2}\int_{0}^{x}K(y,x-y)f(t,y)f(t,x-y)dy\nonumber \\
 & - & f(x)\int_{0}^{\infty}K(x,y)f(y)dy\label{eq:contin agg}
\end{eqnarray}
where $K(x,y)$ is the aggregation kernel indicating the rate at which
aggregates of volumes $x$ and $y$ join together creating an aggregate
of volume $x+y$. Notice the first integral, $A_{in}(f)$, describes
aggregates with volumes $y$ and $x-y$ aggregating to a combined
volume $x$, and the second integral, $A_{out}(f)$, models interactions
between the aggregate of volume $x$ with all other aggregates of
volume $y$ forming an aggregate of volume $x+y$. Also, note that
the aggregation kernel $K(x,y)$ is positive and symmetric
\[
0<K(x,y)=K(y,x),\,(x,y)\in\mathbb{R}_{+}^{2},
\]
as well as homogeneous, which in this field, is defined as
\begin{equation}
K(\lambda x,\lambda y)=\lambda^{m}K(x,y),\,\,\,\lambda>0,\, m\ge0,\, x,y<\infty.\label{eq:homgen def}
\end{equation}
Because only aggregation is considered, the total number of particles
decreases with each coagulation event.

In practice, when we model experimental data, we often find that the
measurements made by experimental devices can produce heteroscedasticity
in the data, i.e., data with non-constant variation. An advantage
of the OGSFs, in both the discrete and continuous versions, lies in
their incorporation of a weighted residual sum of squares (WRSS),
which allows for differing variance. In this context, the OGSFs give
a greater weight to measurements with smaller variation.

Additionally, when we know specifically how the variance differs,
we transform the model to overcome heteroscedasticity. For example,
with many experimental devices, the measurement error grows with the
size of the quantity measured resulting in a log-normal error distribution
(as described in \cite{Carroll1988} and utilized in \cite{BortzEtal2008bmb}).
With a log-normal error distribution, the analog of (\ref{eq:noisy observation})
would be
\[
\log\left[y(t)\right]=\log\left[f(t,\boldsymbol{\theta})\right]+\epsilon\,,
\]
where $\epsilon$ has a normal distribution with zero mean and variance,
$\sigma^{2}$. For the purposes of this paper, we will only consider
constant variance.

To illustrate an application of the PGSFs to the continuous model
in (\ref{eq:contin agg}), we choose three coagulation kernels, the
constant, the additive, and the multiplicative, for which known solutions
to (\ref{eq:smolu}) exist. In Section \ref{sub:Set-up}, we list
the three solutions with proper placement of the constant parameter,
$\alpha$. Additionally, we justify our choice of minimums for $\mathcal{D}$.
Then in Section \ref{sub:Summation-choices}, we discuss the benefits
and drawbacks of different choices for the order of summation when
calculating the PGSFs.

\subsection{Set up\label{sub:Set-up}}

In Table \ref{solutions}, we list the three kernels studied in this
work and the source of the known solution. Note that analytical solutions
presented in the literature commonly assume a constant, $\alpha=1$,
in the aggregation kernels. We aim to identify the value of $\alpha$,
so we incorporate it as the general constant.

\begin{table}[H]
\centering{}%
\begin{tabular}{cccl}
\hline 
 & $K(x,y)$ & $f(t,x;\alpha)$ for general constant,  & Source\tabularnewline
 &  & $\alpha\in\mathbb{R}_{+}<\infty$ & \tabularnewline
\hline 
Constant & $\alpha$ & $f(t,x;\alpha)=\left(\frac{2}{\alpha t}\right)^{2}e^{\frac{-2x}{\alpha t}},$ & Aldous \cite{Aldous1999}\tabularnewline
 &  & $\text{for\,\,}x\in[0,\infty),\alpha t\in(0,\infty)$ & \tabularnewline
Additive & $\alpha(x+y)$ & $\frac{1}{\sqrt{2\pi}}x^{-3/2}\left(e^{-\alpha t}\right)e^{-x/\left(2e^{2\alpha t}\right)},$ & Menon\tabularnewline
 &  & $\text{for}\,\, x\in(0,\infty),\alpha t\in[0,\infty)$ & and Pego \cite{Menon2006}\tabularnewline
Multiplicative & $\alpha xy$ & $\frac{1}{\sqrt{2\pi}}x^{-5/2}e^{-(1-\alpha t)^{2}x/2},$ & Menon\tabularnewline
 &  & $\text{for}\,\, x\in(0,\infty),\alpha t\in[0,1)$ & and Pego \cite{Menon2006}\tabularnewline
\end{tabular}\caption{Solutions to the Smoluchowski coagulation equation}
\label{solutions}
\end{table}

The PGSFs are defined on a domain which starts at a point $\boldsymbol{0}\in\mbox{\ensuremath{\mathbb{R}}}^{N}$.
For our purposes, the PGSFs incorporate $\left(\frac{\partial f}{\partial\alpha}\right)^{2}$,
so when we examine the lower ends of $\mathcal{D}$, we must consider
the limit as $t,x\rightarrow0^{+}$ of $\left(\frac{\partial f}{\partial\alpha}\right)^{2}$.
As an example, consider the constant kernel where
\[
\frac{\partial f}{\partial\alpha}=\frac{8}{t^{2}\alpha^{3}}e^{\frac{-2x}{\alpha t}}\left[\frac{x}{t\alpha}-1\right].
\]
 Choosing the path along $x=0$ demonstrates an infinite limit,
\begin{eqnarray}
\lim_{(t,0)\rightarrow(0^{+},0^{+})}\left(\frac{\partial f}{\partial\alpha}\right)^{2} & = & \lim_{(t,0)\rightarrow(0^{+},0^{+})}\frac{64}{t^{4}\alpha^{6}}e^{\frac{-4x}{\alpha t}}\left[\frac{x}{t\alpha}-1\right]^{2}\nonumber \\
 & = & \lim_{(t,0)\rightarrow(0^{+},0^{+})}\frac{64}{t^{4}\alpha^{6}}.\label{eq:derivsqdsingularity}
\end{eqnarray}
The infinite limit in (\ref{eq:derivsqdsingularity}) helps guide
our choice of $\underbar{\ensuremath{t}}=0.2$ because it ensures
our PGSFs calculations remain within computer precision. Similar analysis
leads to choices of $\underbar{\ensuremath{x}}=0.1$ for both the
additive kernel and the multiplicative kernel. In Appendix \ref{sec:domain-choices},
we present similar calculations in detail for all three kernels, justifying
the choice of the lower bounds in each case.

\subsection{Summation choices for calculating PGSFs\label{sub:Summation-choices}}

We note that in (\ref{eq:2DcontGSF}), one is faced with a choice
of which variable is summed first. This choice is not encountered
in the OGSFs context. To calculate the generalized sensitivity, we
can calculate the numerator in (at least) three ways.

One possible choice, which we designate as the \textit{Simultaneously
Incremental }(SI) method, involves summing along the spatial axis
to $x_{s}$ with $s=1,\dots,N_{x}$ and then incrementing time. For
the SI method,
\begin{eqnarray}
gs_{\mathsmaller{SI}}(t_{k},x_{s};\alpha) & = & \frac{\int_{0}^{t_{k}}\int_{0}^{x_{s}}\left(\frac{\partial f}{\partial\alpha}\right)^{2}dxdt}{gs(\overline{t},\overline{x};\alpha)}\nonumber \\
 & \approx & \frac{\sum_{i=1}^{k}\sum_{j=1}^{s}\left(\frac{\partial f}{\partial\alpha}(t_{i},x_{j};\alpha)\right)^{2}\Delta x_{j}\Delta t_{i}}{gs(\overline{t},\overline{x};\alpha)},\label{eq:SI meth}
\end{eqnarray}
where $\Delta x_{j}=x_{j+1}-x_{j}$ and $\Delta t_{j}=t_{j+1}-t_{j}$,
where $N_{x}$ represents the number of volume points. Unless otherwise
specified, we space our grids uniformly. Note that in (\ref{eq:SI meth}),
the order of summation does not matter.

We designate the second method, the \textit{All Size, Incremental
in Time} (ASIT) method, with which we sum along the entire size-axis
before we increment time. For the ASIT method, we denote $(t_{k},x_{s})=(t,x)_{i}$
such that $i=s+(k-1)N_{x}$ with $k=1,2,\dots,N_{t}$ and $s=1,2,\dots,N_{x}$.
Then for the ASIT method, 
\begin{eqnarray*}
gs_{\mathsmaller{ASIT}}(\left(t,x\right)_{i};\alpha)=gs_{\mathsmaller{ASIT}}(t_{k},x_{s};\alpha) & = & \frac{\int_{0}^{t_{k}}\int_{0}^{x_{s}}\left(\frac{\partial f}{\partial\alpha}\right)^{2}dxdt}{gs(\overline{t},\overline{x};\alpha)}\\
 & \approx & \frac{\sum_{j=1}^{i}\left(\frac{\partial f}{\partial\alpha}(\left(t,x\right)_{j};\alpha)\right)^{2}\Delta x_{j}\Delta t_{j}}{gs(\overline{t},\overline{x};\alpha)}.
\end{eqnarray*}

Lastly, we designate the third method, the \textit{All Time, Incremental
in Size} (ATIS) method, with which we sum along the entire time-axis
before we increment the size dimension. In this case, we denote $(t_{k},x_{s})=(t,x)_{j}$
such that $j=k+(s-1)N_{t}$ with $k=1,2,\dots,N_{t}$ and $s=1,2,\dots,N_{x}$.
Therefore
\begin{eqnarray*}
gs_{\mathsmaller{ATIS}}(\left(t,x\right)_{j};\alpha)=gs_{\mathsmaller{ATIS}}(t_{k},x_{s};\alpha) & = & \frac{\int_{0}^{t_{k}}\int_{0}^{x_{s}}\left(\frac{\partial f}{\partial\alpha}\right)^{2}dxdt}{gs(\overline{t},\overline{x};\alpha)}\\
 & \approx & \frac{\sum_{i=1}^{j}\left(\frac{\partial f}{\partial\alpha}(\left(t,x\right)_{i};\alpha)\right)^{2}\Delta x_{j}\Delta t_{j}}{gs(\overline{t},\overline{x};\alpha)}.
\end{eqnarray*}
For all three methods, we compute the denominator of our generalized
sensitivity, 
\begin{eqnarray}
gs(\overline{t},\overline{x};\alpha) & = & \int_{0}^{\overline{t}}\int_{0}^{\overline{x}}\left(\frac{\partial f}{\partial\alpha}(t,x;\alpha)\right)^{2}dxdt\nonumber \\
 & = & \sum_{i=1}^{N_{t}-1}\int_{t_{i}}^{t_{i+1}}\sum_{j=1}^{N_{x}-1}\int_{x_{j}}^{x_{j+1}}\left(\frac{\partial f}{\partial\alpha}(t,x;\alpha)\right)^{2}dxdt\nonumber \\
 & \approx & \sum_{i=1}^{N_{t}-1}\sum_{j=1}^{N_{x}-1}\left(\frac{\partial f}{\partial\alpha}(t_{i},x_{j};\alpha)\right)^{2}\Delta x_{j}\Delta t_{i}.\label{eq:discgensens}
\end{eqnarray}
In Section \ref{sec:Results}, we offer justification for calculating
the PGSFs via the SI method rather than via either the ASIT or ATIS
methods.

\section{Determining $\mathcal{D}^{*}$ for the Smoluchowski coagulation equation
with PGSFs\label{sec:Results}}

In order to apply the PGSFs concept to the Smoluchowski coagulation
equation, we make several decisions. First, we choose three aggregation
kernels, constant, additive, and multiplicative, for which known solutions
exist. Next we choose $\mathcal{D}$ and the number of points on our
grid. These choices need to provide enough information and enough
resolution to extract a meaningful $\mathcal{D}^{*}$. We provide
the details of the impacts of these choices later in this section.
Finally, in order to compute the PGSFs, we choose the advocate for
one of the three summation orders described in Section \ref{sub:Summation-choices}.

\begin{figure}[H]
\centering{}\subfloat[Contour plot of PGSFs vs.~time and volume which we calculate via
the SI method]{\centering{}\includegraphics[scale=0.5]{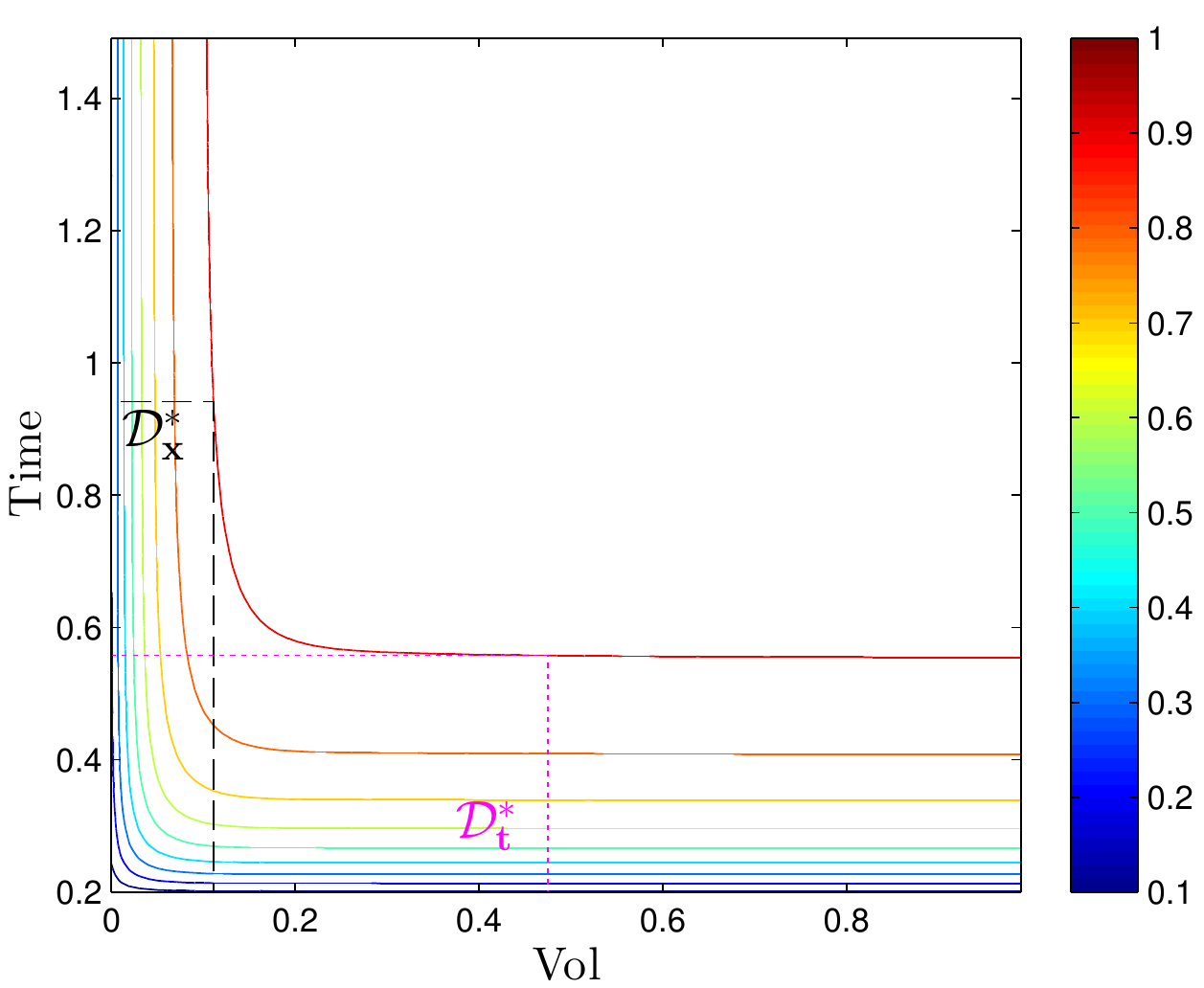}\label{gsfconst}}\hfill{}\subfloat[Contour plot of RIA vs. time and volume]{\centering{}\includegraphics[scale=0.5]{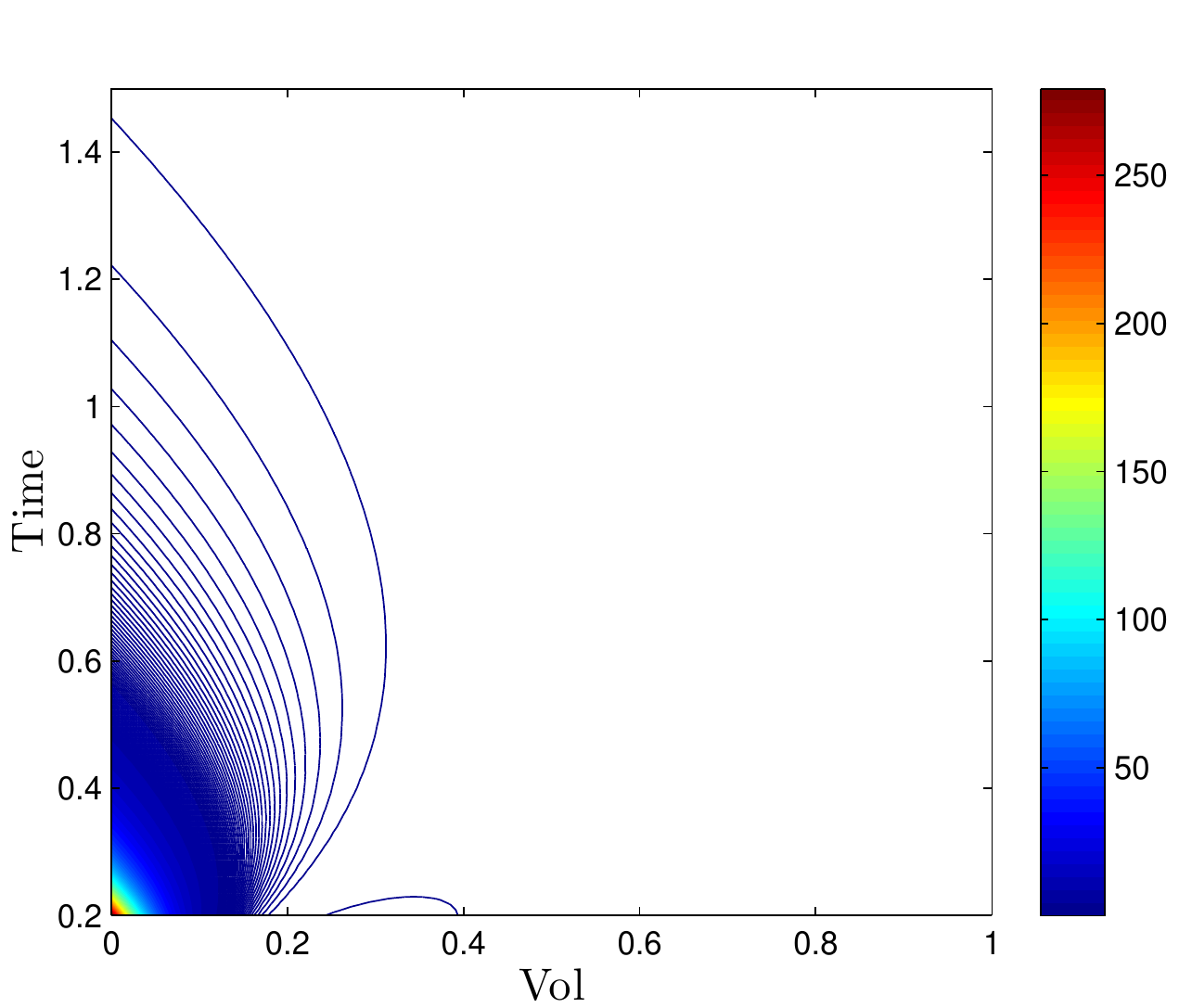}\label{iarconst}}\caption{Generalized Sensitivity and RIA for the constant kernel - subregions
where the largest rates of change occur as the PGSFs transition from
zero to one indicate an approximate $\mathcal{D}^{*}$. The rectangles
in (a) represent optimum subdomains, $\mathcal{D}_{x}^{*}$ and $\mathcal{D}_{t}^{*}$,
as summarized in Table \ref{domaintable}.}
\label{gsfandiarconst}
\end{figure}

\begin{figure}[H]
\centering{}\subfloat[Contour plot of PGSFs vs.~time and volume which we calculate via
the SI method]{\centering{}\includegraphics[scale=0.5]{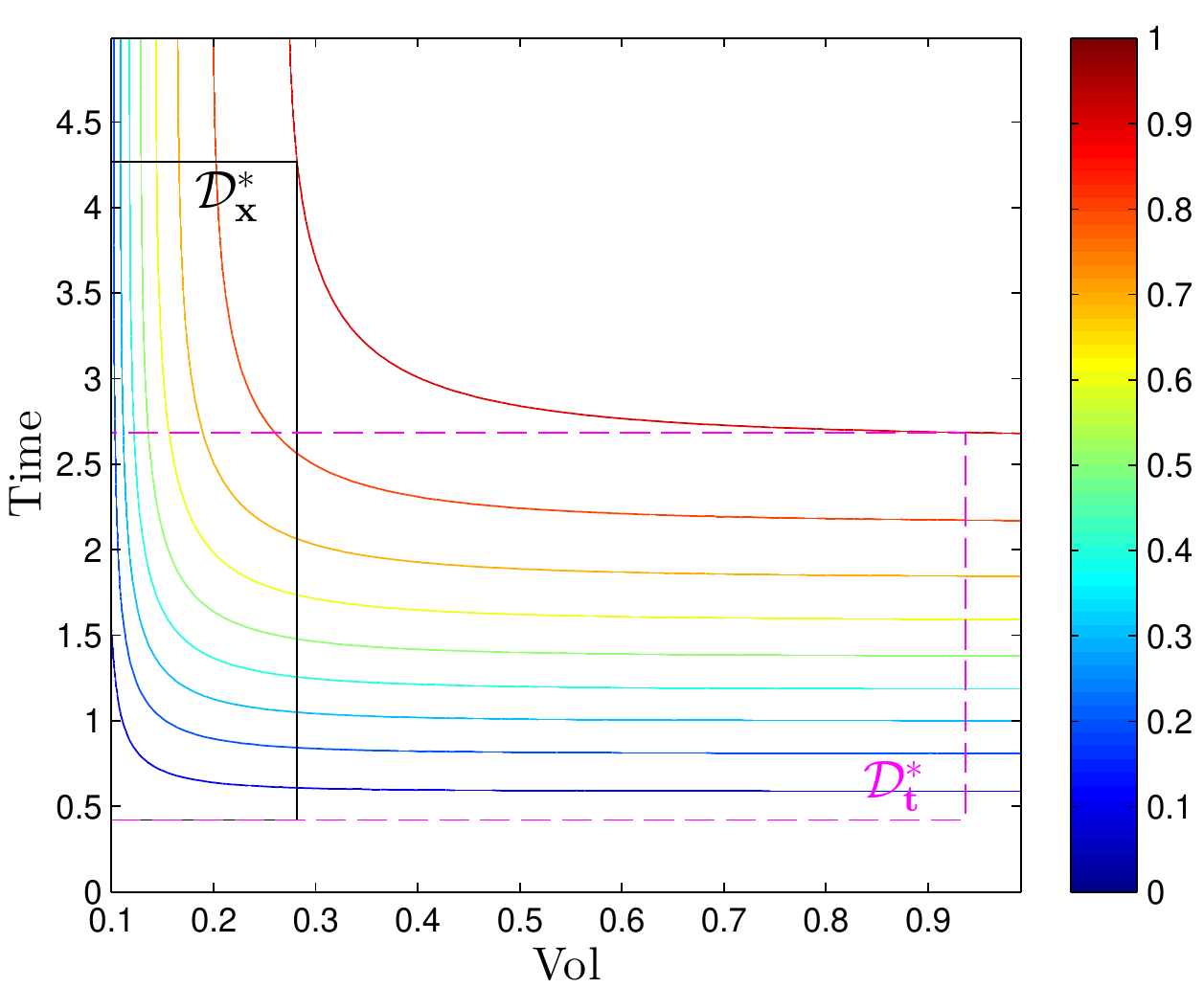}\label{gsfadd}}\hfill{}\subfloat[Contour plot of RIA vs. time and volume]{\centering{}\includegraphics[scale=0.5]{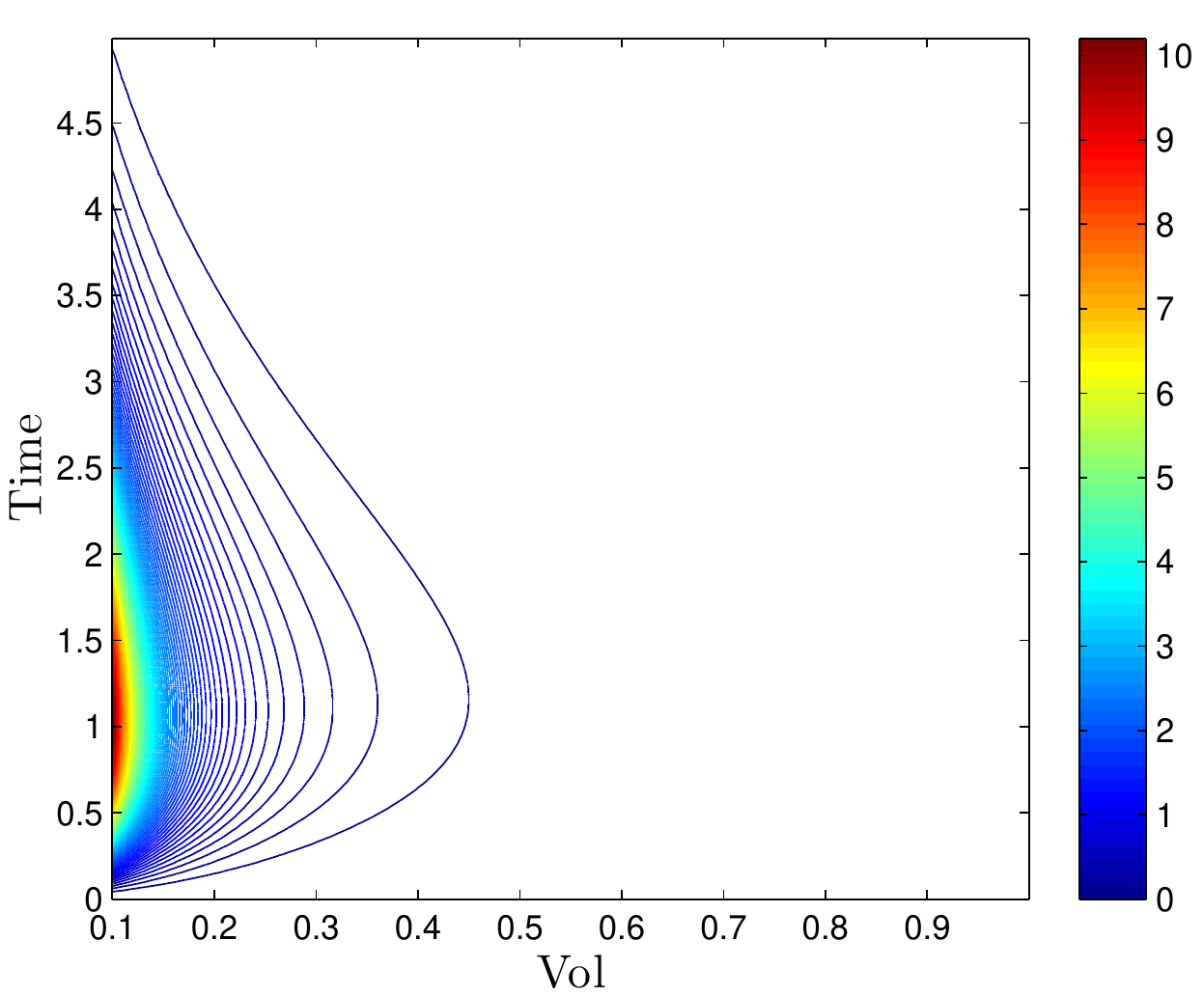}\label{iaradd}}\caption{Generalized Sensitivity and RIA for the additive kernel - subregions
where the largest rates of change occur as the PGSFs transition from
zero to one indicate an approximate $\mathcal{D}^{*}$. The rectangles
in (a) represent optimum subdomains, $\mathcal{D}_{x}^{*}$ and $\mathcal{D}_{t}^{*}$,
as summarized in Table \ref{domaintable}.}
\label{gsfandiaradd}
\end{figure}

\begin{figure}[H]
\centering{}\subfloat[Contour plot of PGSFs vs.~time and volume which we calculate via
the SI method]{\centering{}\includegraphics[scale=0.5]{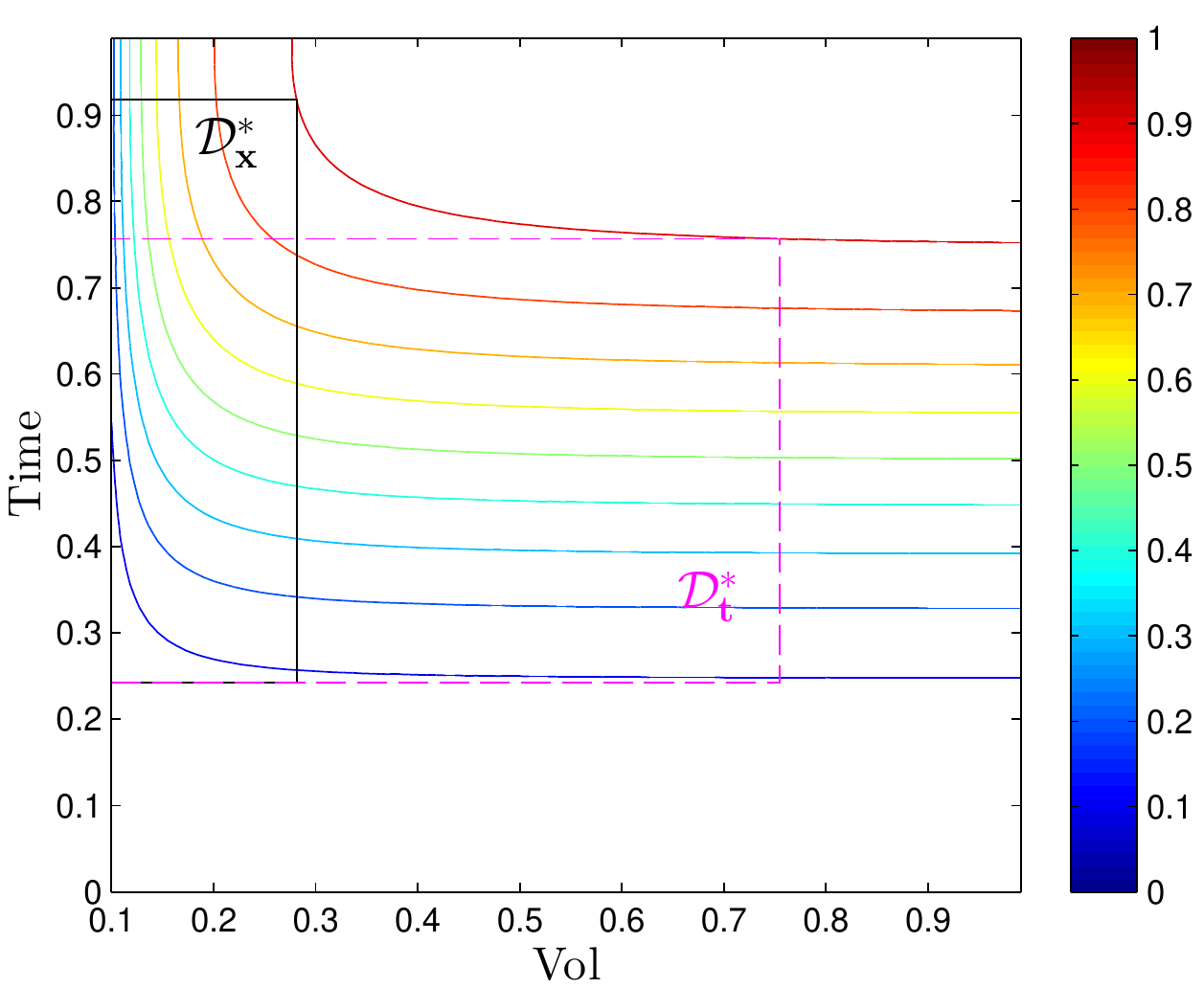}\label{gsfmult}}\hfill{}\subfloat[Contour plot of RIA vs. time and volume]{\centering{}\includegraphics[scale=0.5]{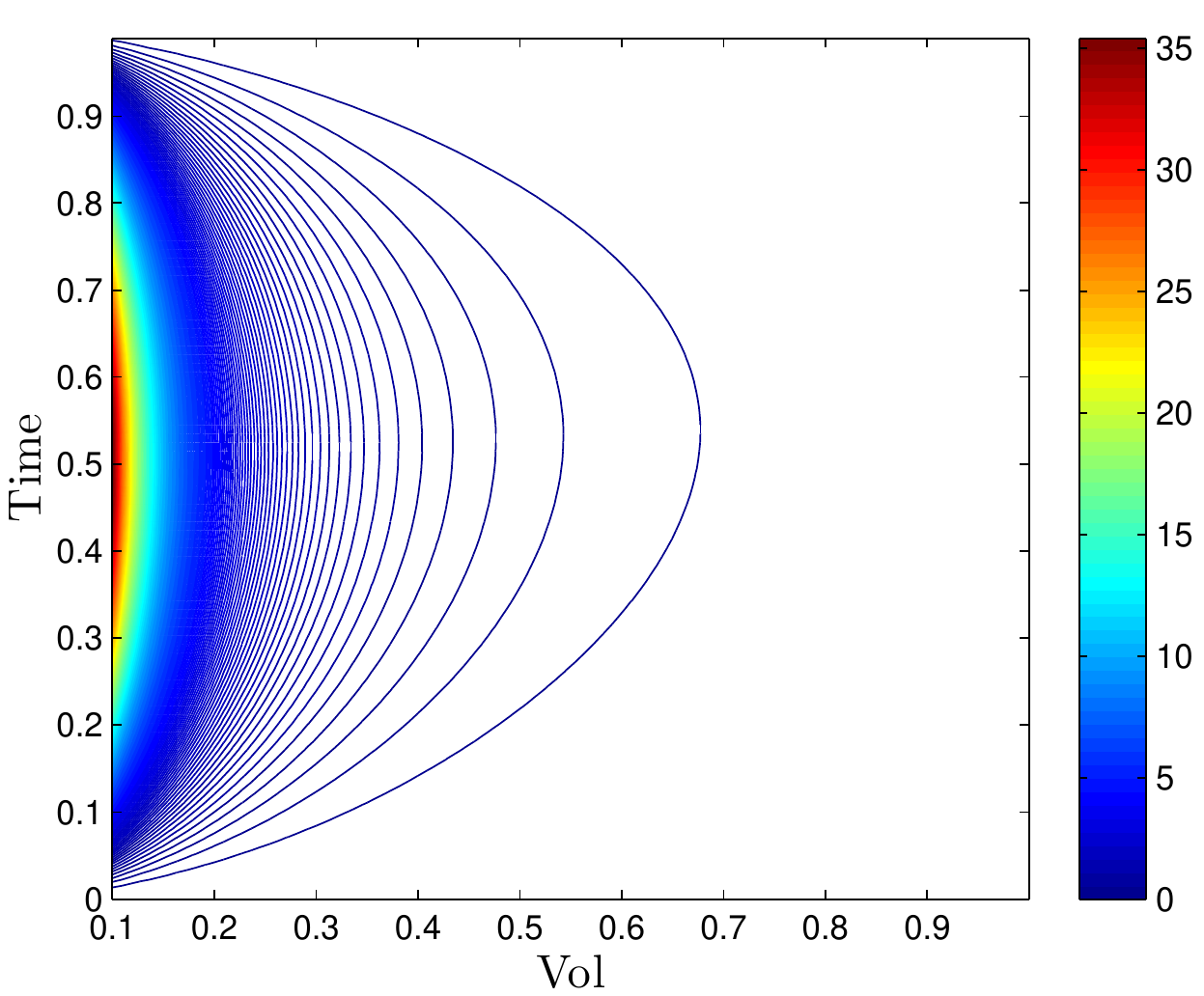}\label{iarmult}}\caption{Generalized Sensitivity and RIA for the multiplicative kernel - subregions
where the largest rates of change occur as the PGSFs transition from
zero to one indicate an approximate $\mathcal{D}^{*}$. The rectangles
in (a) represent optimum subdomains, $\mathcal{D}_{x}^{*}$ and $\mathcal{D}_{t}^{*}$,
as summarized in Table \ref{domaintable}.}
\label{gsfandiarmult}
\end{figure}

In the use of GSFs, a natural question concerns choosing the overall
domain $\mathcal{D}$. For all three kernels, to choose the lower
bounds ($\underbar{\ensuremath{x}}$ and $\underbar{\ensuremath{t}}$)
of $\mathcal{D}$, we face the following dilemma concerning the FTOA.%
\footnote{Recall the FTOA is described in Section \ref{sub:Set-up}. We also
expand upon this issue in Appendix \ref{sec:domain-choices}%
} After we determine $\underbar{\ensuremath{x}}$ and $\underbar{\ensuremath{t}}$,
determining $\overline{x}$ and $\overline{t}$ in conjunction with
grid spacing also presents difficulties. If we space the grid too
widely, the PGSFs reach one on the first step, which does not provide
a meaningful resolution. Furthermore, if we use maximum values for
$\mathcal{D}$ that are too small, we face a potential FTOA as described
in Section \ref{sec:GSF-Theory}. To avoid this artifact, we examine
the PGSFs curves and the RIA curves to ensure that the PGSFs curves
stabilize at one well before the maximum domain limits and to confirm
that the RIA stabilizes near zero. If we do not achieve both of those
criteria, we need to increase $\overline{x}$ or $\overline{t}$ until
we do. For all three kernels that we study, the PGSFs curves in Figures
\ref{gsfconst} , \ref{gsfadd}, and \ref{gsfmult}, do stabilize
at one before $\overline{x}$ and $\overline{t}$, and the RIA stabilizes
near zero in Figures \ref{iarconst}, \ref{iaradd}, and \ref{iarmult}.

\begin{figure}[H]
\centering{}\subfloat[Entire range]{\centering{}\includegraphics[scale=0.5]{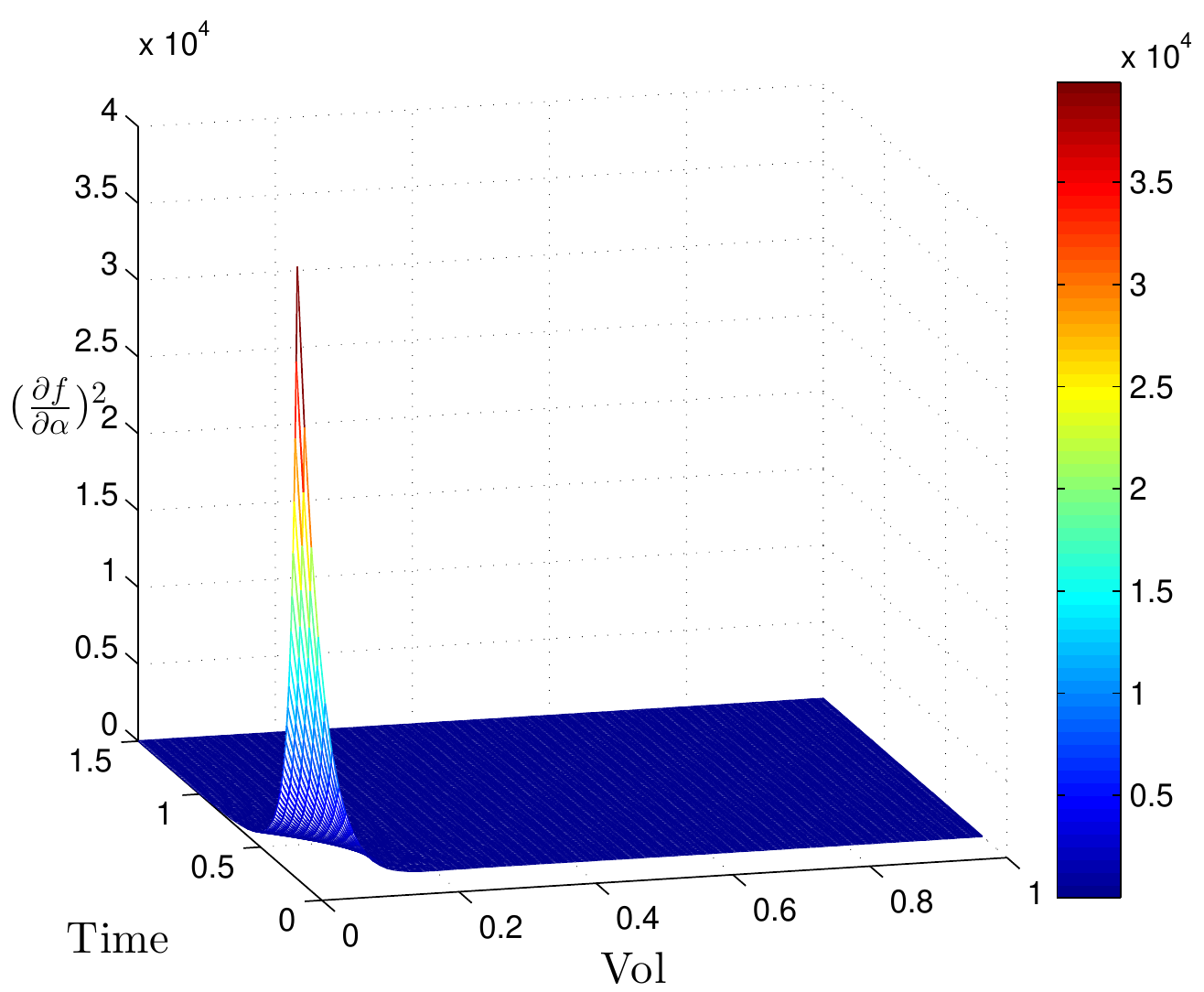}\label{entireplot}}\hfill{}\subfloat[Zoomed in]{\centering{}\includegraphics[scale=0.5]{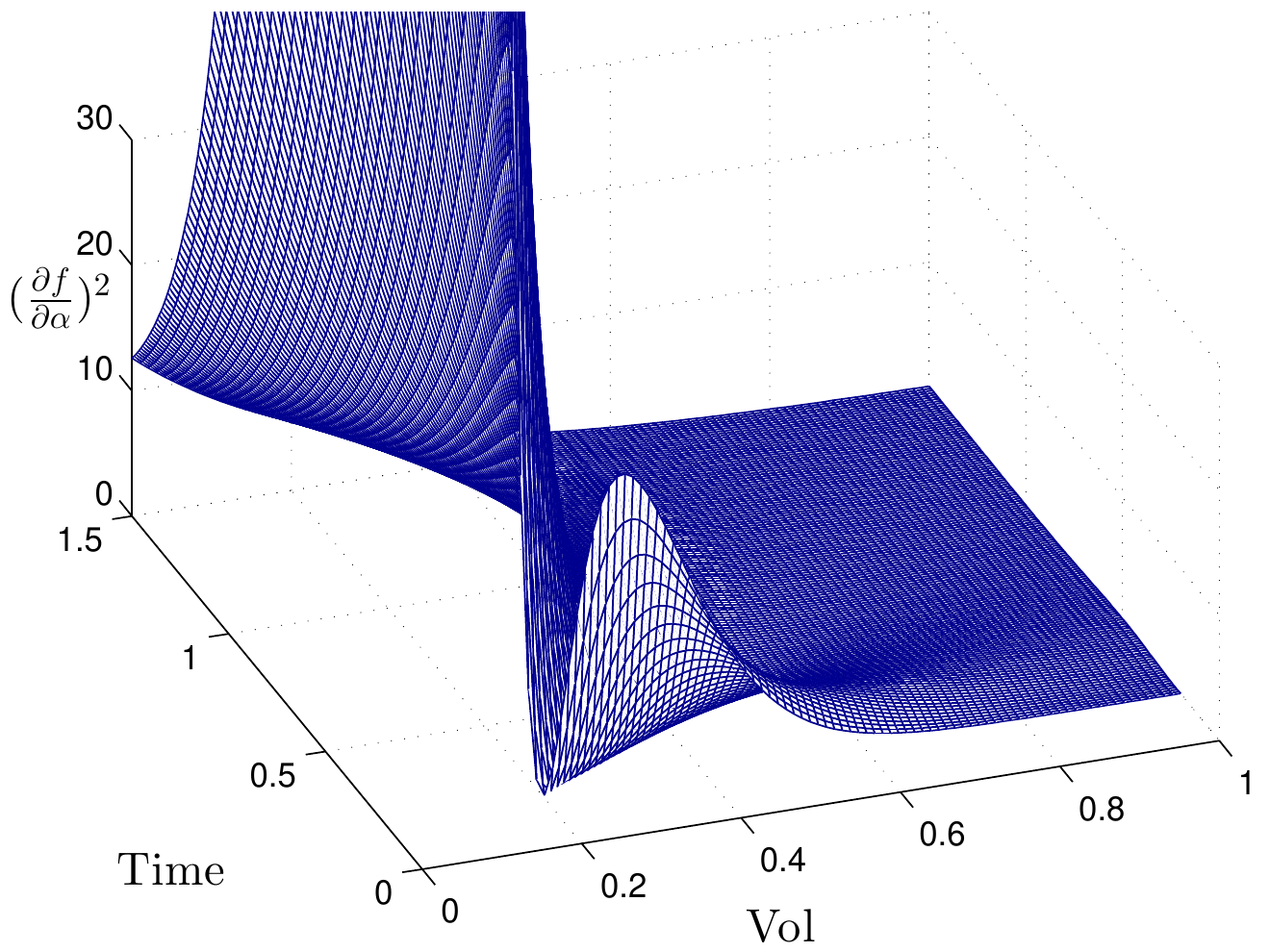}\label{zoomedin}}\caption{$\left(\frac{\partial f(t,x)}{\partial\alpha}\right)^{2}$vs.~$x$
and $t$ for $K(x,y)=\alpha$: In (a) we plot the entire range which
illustrates the spike at small time and small volume points, and in
(b) we illustrate more detail away from the spike.}
\label{derivsqdconstkern}
\end{figure}

The primary purpose of applying PGSFs in our study is to determine
the subdomains, $\mathcal{D}^{*}$, that contain the most important
information relative to estimating the constant, $\alpha$. In our
application of the PGSFs to the Smoluchowski coagulation equation,
we incorporate one parameter, therefore $\left(\frac{\partial f}{\partial\alpha}\right)^{2}$
provides the primary quantity of interest. As depicted in Figure \ref{entireplot},
for the constant kernel, we notice a large spike at small times and
volumes. Then zooming in as depicted in Figure \ref{zoomedin}, we
notice more detail at volumes greater than approximately $0.2$. The
plots in Figure \ref{derivsqdconstkern}, do not clearly indicate
the importance of the subdomain, $x\in[0.2,0.6]$. 
Conversely, the PGSFs and RIA plots allows us to quantify the relative
importance of all the contributions. Figure \ref{gsfandiarconst}
reveals the PGSFs approach one and the rates of information acquisition
approach zero well within $\mathcal{D}$. By implementing the mathematical
strategy in Section \ref{sub:subdomain_determ}, we compute the lower
bounds, $(t_{*},x_{*})=(0.2,0)$, from $\Gamma_{\gamma}$ and the
upper bounds, which range from $(t(x^{*}),x^{*})\approx(0.94,0.11)$
to $(t^{*},x(t^{*}))\approx(0.56,0.48)$, from $\Gamma^{\rho}$. We
achieve these results (and the results for the other two kernels)
by setting $\Delta t=\Delta x=.01$, $\gamma=0.5$, and $\rho=0.1$
and by implementing the SI method described in Section \ref{sub:Summation-choices}.

We can determine $\mathcal{D}^{*}$ for the additive and multiplicative
kernels by performing similar assessments of the PGSFs plots (in Figures
\ref{gsfadd} and \ref{gsfmult}) to ensure we avoid the FTOA. We
then confirm that the rates approach zero in those subdomains in Figures
\ref{iaradd} and \ref{iarmult} respectively. By implementing the
mathematical strategy in Section \ref{sub:subdomain_determ} for the
additive kernel, we compute the lower bounds, $(t_{*},x_{*})=(0.42,0.1)$,
and the upper bounds, which range from $(t(x^{*}),x^{*})\approx(4.27,0.28)$
to $(t^{*},x(t^{*}))\approx(2.69,0.94)$. For the multiplicative kernel,
we compute the lower bounds, $(t_{*},x_{*})=(0.24,0.1)$, and the
upper bounds, which range from $(t(x^{*}),x^{*})\approx(0.920.28)$
to $(t^{*},x(t^{*}))\approx(0.76,0.75)$. We summarize $\mathcal{D}^{*}$
for each aggregation kernel in Table \ref{domaintable}.

{\small }
\begin{table}[H]
\centering{}{\small }%
\begin{tabular}{cccc}
\hline 
{\small $K(x,y)$} & {\small $\mathcal{D}_{x}^{*}=[t_{*},t(x^{*})]\times[x_{*},x^{*}]$} & {\small $\mathcal{D}_{t}^{*}=[t_{*},t^{*}]\times[x_{*},x(t^{*})]$} & {\small Gelation}\tabularnewline
\hline 
{\small $\alpha$} & {\small $[0.2,0.94]\times[0,0.11]$} & {\small $[0.2,0.56]\times[0,0.48]$} & \tabularnewline
{\small $\alpha(x+y)$} & {\small $[0.42,4.27]\times[0.1,0.28]$} & {\small $[0.42,2.69]\times[0.1,0.94]$} & \tabularnewline
{\small $\alpha xy$} & {\small $[0.24,0.92]\times[0.1,0.28]$} & {\small $[0.24,0.76]\times[0.1,0.75]$} & {\small $t=\frac{1}{\alpha}$}\tabularnewline
\end{tabular}{\small \caption{Summary of $\mathcal{D}^{*}$ when estimating the constant in three
aggregation kernels for the Smoluchowski coagulation equation. We
achieve these results by setting $\Delta t=\Delta x=.01$, $\gamma=0.5$,
and $\rho=0.1$ and by implementing the SI method described in Section
\ref{sub:Summation-choices}. The second column reflects an optimum
subdomain when volume data is more costly and the third column denotes
an optimum subdomain when the time data is more costly. Note that
the time subdomain for the multiplicative kernel indicates that the
pertinent information occurs prior to gelation.}
\label{domaintable}}
\end{table}
{\small \par}

The PGSFs for the multiplicative kernel provide another important
result. It is well known that \textit{gelation} occurs for the multiplicative
kernel%
\footnote{With the multiplicative kernel, gelation occurs at $t=\frac{1}{\alpha}.$%
}. When gelation occurs the system experiences growth rapid enough
that aggregates with \textit{infinite} volume develop in finite time
\cite{Wattis2006a}. Mass is not physically lost, but the aggregates
with \textit{infinite} volume possess fundamentally different mathematical
properties than the individual aggregates that make up the gel. We
direct the interested reader to \cite{Ziff1980}, in which Ziff and
Stell provide a thorough description of the implications of various
assumptions on the post-gelation behavior of the solutions and of
the moments. As summarized in Table \ref{domaintable}, our results
provide evidence that the pertinent information necessary for estimating
the constant in $K(x,y)=\alpha xy$ occurs well before gelation. 


\begin{figure}[H]
\centering{}\subfloat[SI Method]{\centering{}\includegraphics[scale=0.35]{gsfconst}}\hfill{}\subfloat[ASIT Method]{\centering{}\includegraphics[scale=0.35]{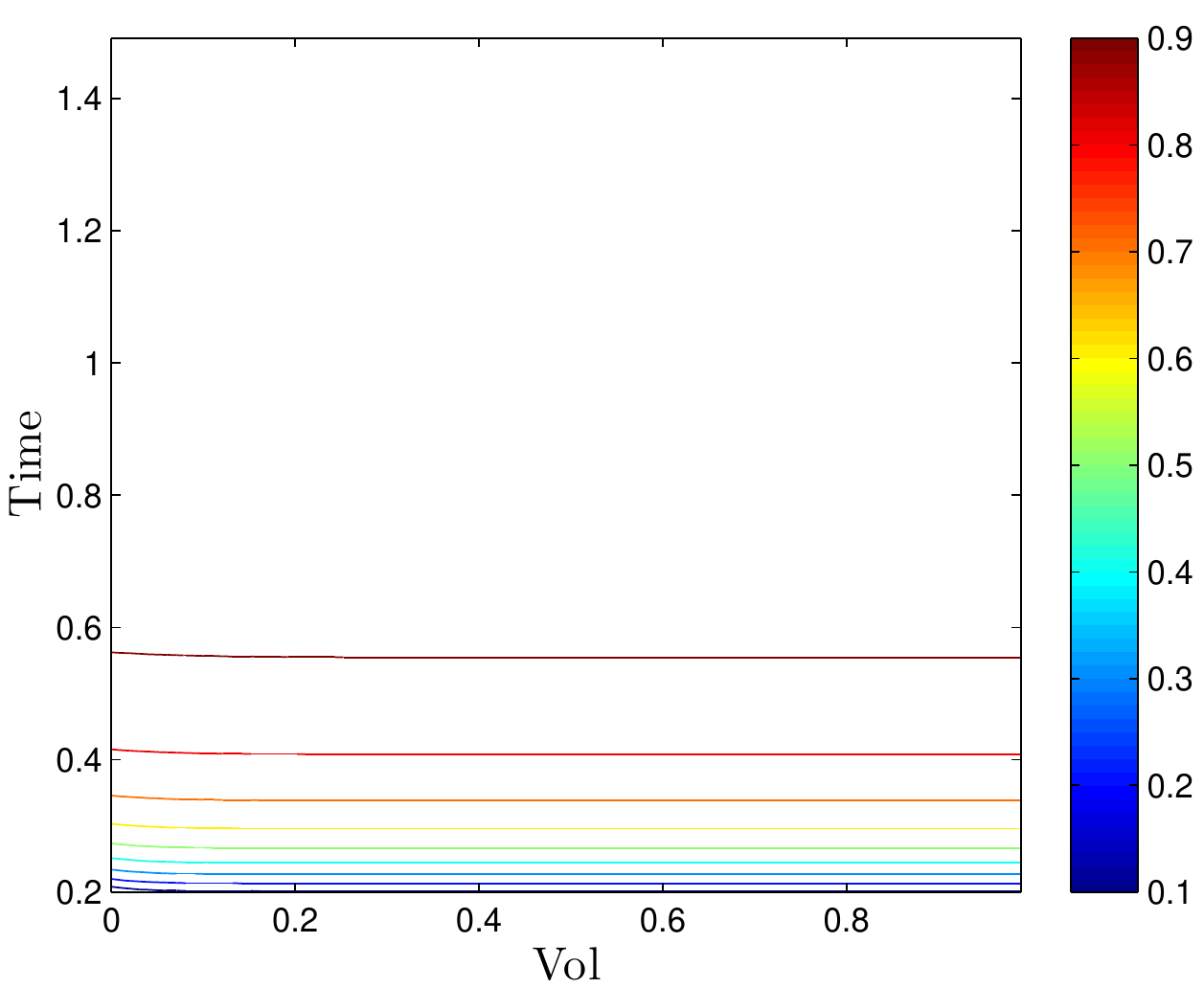}}\hfill{}\subfloat[ATIS Method]{\centering{}\includegraphics[scale=0.35]{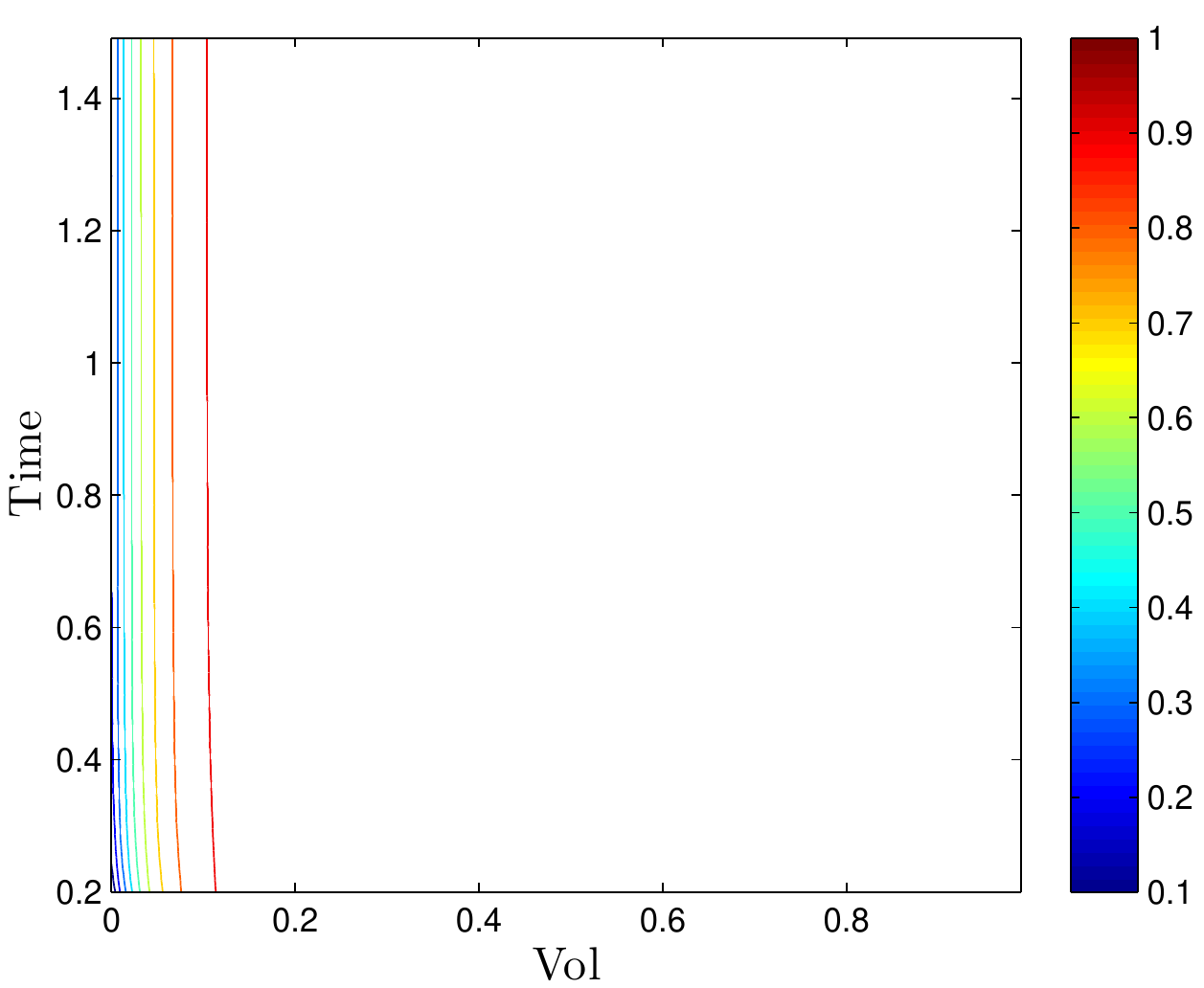}}\caption{Comparison of the the three summation methods for calculating the
PGSFs for the constant kernel: in (a), we sum along $x$ to $x_{s}$
and then increment time. In (b), we sum along $x$ to $x_{N_{x}}$
and then increment time. In (c), we sum along $t$ to $t_{N_{t}}$
and then increment volume. Note that the SI method illustrates a combination
of the ASIT and ATIS methods providing both time and volume indications
of where the generalized sensitivity reaches one on the same plot.}
\label{gsfcomp}
\end{figure}

Finally, as described in Section \ref{sub:Summation-choices}, we
examined three summation methods when calculating the PGSFs. We plot
the constant kernel PGSFs for each of the three methods in Figure
\ref{gsfcomp}. The ASIT method indicates the approximate $\overline{t}$
necessary for the generalized sensitivity to reach one, but it does
not provide an obvious indication of $\overline{x}$. Conversely,
the ATIS method indicates the necessary $\overline{x}$ for the generalized
sensitivity to reach one, but it does not provide useful information
relative to the time domain. However, the SI method simultaneously
illustrates a combination of the ASIT and ATIS methods and provides
both time and volume indications of where the generalized sensitivity
reaches one. We achieve similar results for the additive and multiplicative
kernels. Note that regardless of the summation scheme we use, the
RIA remains the same.

\section{Conclusions and Future Work\label{sec:Conclusions-and-Future}}

In this work, we have extended the concepts of the ODE-based GSFs
introduced by Thomaseth and Cobelli in \cite{Thomaseth1999}, to the
PDE-based GSFs. These PGSFs provide a framework for determining an
optimum subdomain, $\mathcal{D}^{*}$, for size-structured population,
PDE models. We then apply PGSFs to the Smoluchowski coagulation equation,
a popular size-structured population model, to determine $\mathcal{D}^{*}$
for parameter estimation in the constant, additive, and multiplicative
kernels.

To accomplish the goal of determining optimal experimental domains,
we offer a novel mathematical means of determining the entire $\mathcal{D}^{*}$
from generalized sensitivity functions. Specifically for the Smoluchowski
coagulation equation, we determine that pertinent information for
estimating the constant parameter, $\alpha$, occurs in small volume
subdomains. When time data costs less than volume data, we generally
require no larger than $x\approx0.3$. We require no larger than $x\approx0.94$
when time data is more costly than volume data. We also determine
that the most relevant time information occurs early in a coagulation
experiment. How early varies widely among the three kernels with maximum
times ranging from $0.56$ to $4.27$. Our study also acknowledges
the potential for a \emph{force to one arti}\textit{fact}, FTOA, which
is a known weakness of the generalized sensitivity functions. By addressing
this weakness, we determine maxima in $\mathcal{D}$ which eliminate
the artifact. Finally, we also provide results which indicate that
all of the relevant time information for the multiplicative kernel
occurs prior to gelation.

With our application to the Smoluchowski coagulation equation, we
include only one parameter to estimate. Generally, PGSFs allows accounting
for multiple parameters, and in our future work we aspire to study
more sophisticated aggregation kernels which contain multiple parameters.
As is popular in much of the literature, we will examine kernels of
the form, $K(x,y)=\alpha(x^{\mu}y^{\nu}+x^{\nu}y^{\mu})$. 

Additionally, the results generated in Section \ref{sec:Results}
follow from inputting a specific true parameter. Clearly, altering
that parameter could shift $\mathcal{D}^{*}$. As a future step, we
aim to methodically study a range of true parameters and their respective
optimum subdomains.

Lastly, in this work, we study the Smoluchowski coagulation equation,
which models processes involving aggregation only. In the future,
we would like to consider the Smoluchowski coagulation-fragmentation
equation for which we would estimate both aggregation and fragmentation
parameters.

\section{Acknowledgments}

This work was supported in part by the National Science Foundation
grant DMS-1225878. We would also like to thank Dr. John Younger in
the Department of Emergency Medicine at the University of Michigan
for our discussions concerning experimental data.

\bibliographystyle{siam}
\bibliography{mathbioCU}

\appendix

\section{Domain choices for aggregation kernels\label{sec:domain-choices}}

The theoretical $\mathcal{D}$ on which the PGSFs are defined starts
at a point $\boldsymbol{0}\in\mbox{\ensuremath{\mathbb{R}}}^{N}$.
For our purposes, the PGSFs incorporate $\left(\frac{\partial f}{\partial\alpha}\right)^{2}$,
so when we examine the lower ends of $\mathcal{D}$, we consider the
limit as $t,x\rightarrow0^{+}$ of $\left(\frac{\partial f}{\partial\alpha}\right)^{2}$.
In this appendix, we determine $\lim_{t,x\rightarrow\boldsymbol{0}^{+}}\left(\frac{\partial f}{\partial\alpha}\right)^{2}$
for each of the the three coagulation kernels.

For $K(x,y)\equiv\alpha$, 
\[
f(t,x;\alpha)=\left(\frac{2}{\alpha t}\right)^{2}e^{\frac{-2x}{\alpha t}},
\]
therefore
\[
\frac{\partial f}{\partial\alpha}=\frac{8}{t^{2}\alpha^{3}}e^{\frac{-2x}{\alpha t}}\left[\frac{x}{t\alpha}-1\right].
\]
Then 
\[
\left(\frac{\partial f}{\partial\alpha}\right)^{2}=\frac{64}{t^{4}\alpha^{6}}e^{\frac{-4x}{\alpha t}}\left[\frac{x}{t\alpha}-1\right]^{2},
\]
therefore
\[
\lim_{(t,x)\rightarrow(0^{+},0^{+})}\left(\frac{\partial f}{\partial\alpha}\right)^{2}=\lim_{(t,x)\rightarrow(0,0)}\frac{64}{t^{4}\alpha^{6}}e^{\frac{-4x}{\alpha t}}\left[\frac{x}{t\alpha}-1\right]^{2},
\]
which does not exist. Choosing the path $x=0$ demonstrates the infinite
limit,
\begin{eqnarray}
\lim_{(t,0)\rightarrow(0^{+},0^{+})}\left(\frac{\partial f}{\partial\alpha}\right)^{2} & = & \lim_{(t,0)\rightarrow(0^{+},0^{+})}\frac{64}{t^{4}\alpha^{6}}e^{\frac{-4x}{\alpha t}}\left[\frac{x}{t\alpha}-1\right]^{2}\nonumber \\
 & = & \lim_{(t,0)\rightarrow(0^{+},0^{+})}\frac{64}{t^{4}\alpha^{6}}.\label{eq:derivsqdsingularity-1}
\end{eqnarray}
The infinite limit in (\ref{eq:derivsqdsingularity-1}) helps guide
our choice of $\underbar{\ensuremath{t}}=0.2$ for $\mathcal{D}$,
which ensures our PGSFs calculations remain within computer precision. 

For $K(x,y)\equiv\alpha(x+y)$, $\alpha\in\mathbb{R}_{+}<\infty$,
$x\in(0,\infty)$, $t\in[0,\infty$) and as adapted from \cite{Menon2006},
\begin{equation}
f(t,x;\alpha)=\frac{1}{\sqrt{2\pi}}x^{-3/2}\left(e^{-\alpha t}\right)e^{-x/\left(2e^{2\alpha t}\right)},\label{eq:anlytsoladdkern}
\end{equation}
where it naturally follows that
\[
f(0,x;\alpha)=\frac{1}{\sqrt{2\pi}}x^{-3/2}e^{-x/2}.
\]
As we choose $\mathcal{D}$ for the additive kernel, we again consider
\[
\lim_{(t,0)\rightarrow(0^{+},0^{+})}\left(\frac{\partial f}{\partial\alpha}\right)^{2}
\]
and note
\[
\left(\frac{\partial f}{\partial\alpha}\right)^{2}=\frac{t^{2}e^{-2\alpha t}e^{\left(-xe^{-2t}\right)}}{2\pi x^{3}}.
\]
Then for any $t=a$, where $a$ is constant strictly greater than
zero, 
\[
\lim_{x\rightarrow0^{+}}\left(\frac{\partial f}{\partial\alpha}\right)^{2}=\lim_{x\rightarrow0^{+}}\frac{a^{2}e^{-2\alpha a}e^{\left(-xe^{-2a}\right)}}{2\pi x^{3}},
\]
which is infinite. Choosing $\underbar{x}=0.1$ as our minimum value
ensures our PGSFs calculations remain within computer precision.

For $K(x,y)\equiv\alpha xy$, $\alpha\in\mathbb{R}_{+}<\infty$, $x\in(0,\infty)$,
$t\in[0,1$) and as adapted from \cite{Menon2006},
\begin{equation}
f(t,x;\alpha)=\frac{1}{\sqrt{2\pi}}x^{-5/2}e^{-(1-\alpha t)^{2}x/2},\label{eq:anlytsolmultkern}
\end{equation}
where it naturally follows that
\[
f(0,x;\alpha)=\frac{1}{\sqrt{2\pi}}x^{-5/2}e^{-x/2}.
\]
As we choose $\mathcal{D}$ for the multiplicative kernel, we again
consider
\[
\lim_{(t,0)\rightarrow(0^{+},0^{+})}\left(\frac{\partial f}{\partial\alpha}\right)^{2},
\]
and note
\[
\left(\frac{\partial f}{\partial\alpha}\right)^{2}=\frac{t^{2}e^{\left(-x(\alpha t-1)^{2}\right)}(\alpha t-1)^{2}}{2\pi x^{3}}.
\]
Then for any $t=a$, where $a\in(0,1)$ is a constant, 
\[
\lim_{x\rightarrow0^{+}}\left(\frac{\partial f}{\partial\alpha}\right)^{2}=\lim_{x\rightarrow0^{+}}\frac{a^{2}e^{\left(-x(\alpha a-1)^{2}\right)}(\alpha a-1)^{2}}{2\pi x^{3}},
\]
which is infinite. Choosing $\underbar{x}=0.1$ as our minimum value
ensures our PGSFs calculations remain within computer precision.
\end{document}